# Ground States of Coupled Nonlinear Oscillator Systems

Uri Levy

Weizmann Institute of Science, Rehovot 7610001, Israel
E-mail: uri.levy@weizmann.ac.il

## References

**[1]** Feynman, R., R. Leighton, and M. Sands, "The feynman lectures on physics, vol. iii: the new millennium edition: quantum mechanics."

**[2]** Auerbach, Assa, *Interacting electrons and quantum magnetism*. Springer Science & Business Media, 2012.

**[3]** Eilbeck J. C. and Johansson M., 2003, *The discrete nonlinear Schrödinger equation - 20 years on*, Proc. 3rd Conf.: Localization and Energy Transfer in Nonlinear Systems (June 17–21 2002 World Scientific) San Lorenzo de El Escorial, Madrid p 44 (arXiv: nlin/0211049).

**[4]** Meier, J., G. I. Stegeman, D. N. Christodoulides, Y. Silberberg, R. Morandotti, H. Yang, G. Salamo, M. Sorel, and J. S. Aitchison, *Experimental observation of discrete modulational instability*, Physical review letters 92, no. 16 (2004): 163902.

**[5]** Davydov, A. S., and N. I. Kislukha, *Solitary excitons in one-dimensional molecular chains,* physica status solidi (b) 59, no. 2 (1973): 465-470.

**[6]** Davydov, A. S., *The theory of contraction of proteins under their excitation,* Journal of Theoretical Biology 38, no. 3 (1973): 559-569.

**[7]** Davydov, A. S., and N. I. Kislukha, *Solitons in One-Dimensional Molecular Chains*, physica status solidi (b) 75, no. 2 (1976): 735-742.

**[8]** Davydov, Alexander S., *Solitons and energy transfer along protein molecules,* Journal of Theoretical Biology 66, no. 2 (1977): 379-387.

**[9]** Jallouli, Aymen, Najib Kacem, and Noureddine Bouhaddi, *Stabilization of solitons in coupled nonlinear pendulums with simultaneous external and parametric excitations*, Communications in Nonlinear Science and Numerical Simulation 42 (2017): 1-11.

**[10]** Sato M, Hubbard BE, Sievers AJ., *Colloquium: Nonlinear energy localization and its manipulation in micromechanical oscillator arrays.* Reviews of Modern Physics. 2006 Jan 26;78(1):137.

**[11]** Eisenberg, H. S., Y. Silberberg, R. Morandotti, A. R. Boyd, and J. S. Aitchison, *Discrete spatial optical solitons in waveguide arrays,* Physical review letters 81, no. 16 (1998): 3383.

**[12]** Silberberg, Yaron, Yoav Lahini, Yaron Bromberg, Eran Small, and Roberto Morandotti, *Universal correlations in a nonlinear periodic 1D system,* Physical review letters 102, no. 23 (2009): 233904.

**[13]** Polkovnikov, Anatoli, Subir Sachdev, and S. M. Girvin, *Nonequilibrium Gross-Pitaevskii dynamics of boson lattice models,* Physical Review A 66, no. 5 (2002): 053607.

**[14]** Kevrekidis, Panayotis G., *The discrete nonlinear Schrödinger equation: mathematical analysis, numerical computations and physical perspectives,* Vol. 232. Springer Science & Business Media, 2009.

**[15]** Pelinovsky, Dmitry E., *Localization in periodic potentials: from Schrödinger operators to the Gross–Pitaevskii equation*, Vol. 390. Cambridge University Press, 2011.

**[16]** J.C. Eilbeck, P.S. Lomdahl, and A.C. Scott, Physica 16D (1985) 318-338.

**[17]** Rasmussen, K. Ø., T. Cretegny, Panayotis G. Kevrekidis, and Niels Grønbech-Jensen, *Statistical mechanics of a discrete nonlinear system,* Physical review letters 84, no. 17 (2000): 3740.

**[18]** Lederer, Falk, George I. Stegeman, Demetri N. Christodoulides, Gaetano Assanto, Moti Segev, and Yaron Silberberg, *Discrete solitons in optics,* Physics Reports 463, no. 1-3 (2008): 1-126.

**[19]** Rumpf, Benno, *Transition behavior of the discrete nonlinear Schrödinger equation,* Physical Review E 77, no. 3 (2008): 036606.

**[20]** Pitaevskii, Lev Petrovitch, Stringari, Sandro, *Bose-Einstein Condensation and Superfluidity*, International series of monographs on physics 164, Oxford University Press, (2016).

**[21]** B. Rumpf, *Stable and metastable states and the formation and destruction of breathers in the discrete nonlinear Schrödinger equation*, Physica D (Amsterdam) 238D, 2067 (2009).

**[22]** Spohn, Herbert, *Fluctuating hydrodynamics approach to equilibrium time correlations for anharmonic chains,* In Thermal Transport in Low Dimensions, pp. 107-158. Springer, Cham, 2016.

**[23]** Mendl, Christian B., and Herbert Spohn, *Low temperature dynamics of the one-dimensional discrete nonlinear Schrödinger equation.* Journal of Statistical Mechanics: Theory and Experiment, no. 8 (2015): P08028.

**[24]** Smerzi, A., A. Trombettoni, P. G. Kevrekidis, and A. R. Bishop, *Dynamical superfluid-insulator transition in a chain of weakly coupled Bose-Einstein condensates.* Physical review letters 89, no. 17 (2002): 170402.

**[25]** Levy, U., *Distance-Independent or Periodic Solutions of the Discrete Nonlinear Schrödinger Equation*, (2016), unpublished.

**[26]** Kivshar, Yuri S., and Michel Peyrard, *Modulational instabilities in discrete lattices,* Physical Review A 46, no. 6 (1992): 3198.

**[27]** Kivshar, Yuri S., *Self-localization in arrays of defocusing waveguides,* Optics letters 18, no. 14 (1993): 1147-1149.

**[28]** Malomed, Boris, and Michael I. Weinstein, *Soliton dynamics in the discrete nonlinear Schrödinger equation,* Physics Letters A 220, no. 1-3 (1996): 91-96.

- **[29]** Weinstein, Michael I., *Excitation thresholds for nonlinear localized modes on lattices,* Nonlinearity 12, no. 3 (1999): 673.

- **[30]** Greiner, Markus, Olaf Mandel, Tilman Esslinger, Theodor W. Hänsch, and Immanuel Bloch, *Quantum phase transition from a superfluid to a Mott insulator in a gas of ultracold atoms,* nature 415, no. 6867 (2002): 39-44.

- **[31]** Levy U., *Evolution of the phase angle in a nonlinear two-site system*, September 2017, unpublished.

- **[32]** Rebuzzini, Laura, Roberto Artuso, Shmuel Fishman, and Italo Guarneri, *Effects of atomic interactions on quantum accelerator modes*, Physical Review A 76, no. 3 (2007): 031603.

- **[33]** Roberts, B. 2020. Time Reversal. *The Routledge Companion to the Philosophy of Physics*, Eleanor Knox and Alistair Wilson (eds).

- **[34]** Mithun, Thudiyangal, Yagmur Kati, Carlo Danieli, and Sergej Flach, *Weakly nonergodic dynamics in the Gross-Pitaevskii lattice,* Physical review letters 120, no. 18 (2018): 184101.

- **[35]** Levy, Uri, and Yaron Silberberg, *Equilibrium temperatures of discrete nonlinear systems*, Physical Review *B* 98, no. 6 (2018): 060303.

- **[36]** Flach, Sergej, and Andrey V. Gorbach, *Discrete breathers—advances in theory and applications,* Physics Reports 467, no. 1-3 (2008): 1-116.

- **[37]** Levy, Uri, Ken Yang, Noam Matzliah, and Yaron Silberberg, *Universal correlations after thermalization in periodic nonlinear systems,* Journal of Physics B: Atomic, Molecular and Optical Physics 51, no. 3 (2018): 035401.

- **[38]** G. Mussardo, *Statistical field theory: an introduction to exactly solved models in statistical physics*, Oxford University Press, 2010).

- **[39]** Colcelli, Andrea, Giuseppe Mussardo, and Andrea Trombettoni, *Deviations from off-diagonal long-range order in one-dimensional quantum systems,* EPL (Europhysics Letters) 122, no. 5 (2018): 50006.

- **[40]** Esteve, J., C. Gross, A. Weller, S. Giovanazzi, and M. K. Oberthaler, *Squeezing and entanglement in a Bose–Einstein condensate,* Nature 455, no. 7217 (2008): 1216-1219.

- **[41]** Huber, Sebastian D., Barbara Theiler, E. Altman, and Gianni Blatter, *Amplitude mode in the quantum phase model,* Physical review letters 100, no. 5 (2008): 050404.

- **[42]** Hermann Claudine, *Statistical Physics: including applications to condensed matter,* Springer Science & Business Media, 2006.

- **[43]** U. Levy, *Equilibrium Temperatures of Discrete Systems at Weak Nonlinearities*, Sept. 2018, unpublished.

- **[44]** Yariv, Amnon, *Coupled-mode theory for guided-wave optics*, IEEE Journal of Quantum Electronics 9, no. 9 (1973): 919-933.

- **[45]** Lomdahl, Peter S., *Nonlinear dynamics of globular proteins*, In Nonlinear electrodynamics in biological systems, pp. 143-154. Springer, Boston, MA, 1984.

[46] Scott, A. C., and L. Macneil, *Binding energy versus nonlinearity for a "small" stationary soliton*, Physics Letters A 98, no. 3 (1983): 87-88.

[47] Kolovsky, Andrey R., *Treating many-body quantum systems by means of classical mechanics*, In Emergent Complexity from Nonlinearity, in Physics, Engineering and the Life Sciences, pp. 37-48. Springer, Cham, 2017.

[48] Christodoulides, D. N., and R. I. Joseph, *Discrete self-focusing in nonlinear arrays of coupled waveguides*, Optics letters 13, no. 9 (1988): 794-796.

[49] Blundell Stephen, *Magnetism in Condensed Matter*, Oxford Master Series in Physics, Oxford U.P., New York, 2001, 1st Edition

[50] Betti, Silvello, Elisa Duca, Mauro Giaconi, and Giorgia Parca, *Modulation instability and conservation of energy: Toward a new model*, Microwave and Optical Technology Letters 53, no. 10 (2011): 2411-2414.

[51] Archambault, P., S. T. Bramwell, J-Y. Fortin, P. C. W. Holdsworth, S. Peysson, and J-F. Pinton, *Universal magnetic fluctuations in the two-dimensional XY model*, Journal of applied physics 83, no. 11 (1998): 7234-7236.

[52] Ramos, Alba, Lucas Fernández-Alcázar, Tsampikos Kottos, and Boris Shapiro, *Optical Phase Transitions in Photonic Networks: a Spin-System Formulation*, Physical Review X10, no. 3 (2020): 031024.

[53] Garcia-Ripoll, Juan J., Miguel A. Martin-Delgado, and J. Ignacio Cirac, *Implementation of spin Hamiltonians in optical lattices*,. Physical review letters 93, no. 25 (2004): 250405.

[54] Kennett, Malcolm P., *Out-of-equilibrium dynamics of the Bose-Hubbard model*, ISRN Condensed Matter Physics 2013 (2013).

[55] Garanovich, Ivan L., Stefano Longhi, Andrey A. Sukhorukov, and Yuri S. Kivshar, *Light propagation and localization in modulated photonic lattices and waveguides*, Physics Reports 518, no. 1-2 (2012): 1-79.

[56] Levy, Uri, and Yaron Silberberg, *Electrical-Field Distributions in Waveguide Arrays - Exact and Approximate*, arXiv preprint arXiv:1401.0642 (2014).

[57] Levy, Uri, Stanislav Derevyanko, and Yaron Silberberg, *Light modes of free space,* In Progress in Optics, vol. 61, pp. 237-281. Elsevier, 2016.

[58] Levy, Uri, Yaron Silberberg, and Nir Davidson, *Mathematics of vectorial Gaussian beams*, Advances in Optics and Photonics 11, no. 4 (2019): 828-891.

[59] Thompson, Ian, Mathematical methods for Bloch waves and Rayleigh–Bloch waves, Loughborough University."

[60] Yeh, Pochi, Amnon Yariv, and Chi-Shain Hong, *Electromagnetic propagation in periodic stratified media. I. General theory*, JOSA 67, no. 4 (1977): 423-438.

[61] Mandelik, Daniel, H. S. Eisenberg, Y. Silberberg, R. Morandotti, and J. S. Aitchison, *Band-gap structure of waveguide arrays and excitation of Floquet-Bloch solitons*, Physical review letters 90, no. 5 (2003): 053902.

[62] Orefice, A., R. Giovanelli, and D. Ditto, *Complete Hamiltonian Description of wave-like features in Classical and Quantum Physics,* Foundations of Physics 39, no. 3 (2009): 256.

- **[63]** Marte, Monika AM, and Stig Stenholm, *Paraxial light and atom optics: the optical Schrödinger equation and beyond*, Physical Review A 56, no. 4 (1997): 2940.

- **[64]** A. Yariv, *Optical Electronics* (Saunders College Publishing, Philadelphia, 1991), 4th ed., pp. 519 – 524.

- **[65]** Boyd, Robert W., Nonlinear optics, Academic press, 2019.

- **[66]** Rakuljic, George A., Amnon Yariv, and Ratnakar Neurgaonkar, *Photorefractive Properties of Undoped and Doped Single Crystal SBN: 60,* In Advances in Materials for Active Optics, vol. 567, pp. 18-25. International Society for Optics and Photonics, 1986.

- **[67]** Efremidis, Nikos K., Suzanne Sears, Demetrios N. Christodoulides, Jason W. Fleischer, and Mordechai Segev, *Discrete solitons in photorefractive optically induced photonic lattices,* Physical Review E 66, no. 4 (2002): 046602.

- **[68]** Fleischer Jason W., Tal Carmon, Mordechai Segev, Nikos K. Efremidis, and Demetrios N. Christodoulides, *Observation of discrete solitons in optically induced real time waveguide arrays,* Physical review letters 90, no. 2 (2003): 023902.

- **[69]** Chen Feng, Milutin Stepić, Christian E. Rüter, Daniel Runde, Detlef Kip, Vladimir Shandarov, Ofer Manela, and Mordechai Segev, *Discrete diffraction and spatial gap solitons in photovoltaic LiNbO3 waveguide arrays*, Optics express 13, no. 11 (2005): 4314-4324.

- **[70]** Frostig Hadas, Itamar Vidal, Robert Fischer, Hanan Herzig Sheinfux, and Yaron Silberberg, *Observation of rogue events in non-Markovian light*, arXiv preprint arXiv:2002.12180 (2020).

- **[71]** Assanto, Gaetano, and Marco Peccianti, *Spatial solitons in nematic liquid crystals*, IEEE journal of quantum electronics 39, no. 1 (2003): 13-21.

- **[72]** Fratalocchi, Andrea, Gaetano Assanto, Kasia A. Brzdąkiewicz, and Mirek A. Karpierz, *Discrete light propagation and self-trapping in liquid crystals*, Optics express 13, no. 6 (2005): 1808-1815.

- **[73]** Matuszewski, Michal, Christian R. Rosberg, Dragomir N. Neshev, Andrey A. Sukhorukov, Arnan Mitchell, Marek Trippenbach, Michael W. Austin, Wieslaw Krolikowski, and Yuri S. Kivshar, *Crossover from self-defocusing to discrete trapping in nonlinear waveguide arrays*, Optics express 14, no. 1 (2006): 254-259.

- **[74]** Swartzlander Jr, G. A., D. R. Andersen, J. J. Regan, H. Yin, and A. E. Kaplan, *Spatial dark-soliton stripes and grids in self-defocusing materials*, Physical review letters 66, no. 12 (1991): 1583.

- **[75]** Morandotti, R., H. S. Eisenberg, Y. Silberberg, M. Sorel, and J. S. Aitchison, *Self-focusing and defocusing in waveguide arrays*, Physical review letters 86, no. 15 (2001): 3296.

- **[76]** Jaksch, Dieter, Christoph Bruder, Juan Ignacio Cirac, Crispin W. Gardiner, and Peter Zoller, *Cold bosonic atoms in optical lattices*, Physical Review Letters 81, no. 15 (1998): 3108.

**[77]** Lewenstein, Maciej, Anna Sanpera, and Veronica Ahufinger, *Ultracold Atoms in Optical Lattices: Simulating quantum many-body systems*. Oxford University Press, 2012.

**[78]** Struck, Julian, Christoph Ölschläger, Malte Weinberg, Philipp Hauke, Juliette Simonet, André Eckardt, Maciej Lewenstein, Klaus Sengstock, and Patrick Windpassinger, *Tunable gauge potential for neutral and spinless particles in driven optical lattices*, Physical review letters 108, no. 22 (2012): 225304.

**Abstract**

The dynamics of coupled nonlinear oscillator systems is often described by the classical discrete nonlinear Schrödinger equation (DNLSE). In its simplest version, the DNLSE is made up of two terms – a nearest-neighbor hopping term and an on-site cubic nonlinear term. Each of the terms is preceded by a coefficient that can take on either a positive or a negative sign. Each of the DNLSE versions is derived from a corresponding equivalent Hamiltonian. The result is a small family of four versions of the DNLSE Hamiltonian, each with its own associated ground state, all indeed scattered in myriad of scientific publications.

Here we present a comprehensive picture for the ground states of DNLSE systems, summarize existing results and provide new insights.

First we classify the four DNLSE Hamiltonians into two pairs according to the sign of the nonlinear term – a "positive/negative Hamiltonian pair" if the sign of the nonlinear term is positive/negative respectively. Ground states of the positive Hamiltonian pair are *discrete plane waves* in either a ferromagnetic-like or an antiferromagnetic-like configuration, depending on the sign of the hopping term. Ground states of the negative Hamiltonian pair are either unstaggered or staggered *site-centered discrete breathers*.

The instantaneous state of a DNLSE system is described by a set of one-parameter complex functions each with its own amplitude and phase. We show that except for the sign of the phase, a ground state associated with a positive/negative Hamiltonian is the *maximum* energy state associated with the sign-reversed (negative/positive) Hamiltonian.

Next we discuss some properties of the ground states associated with the positive-Hamiltonian pair - entropy, temperature, correlations and stability. We extend our ground state stability discussion to include excited plane waves. We propose to engineer a specific perturbation that preserves both density and energy – the two conserved quantities of a DNLSE system – and to test plane wave's stability based on entropy change. We show that under such conserved-quantities-preserved perturbation, all excited plane waves are entropy-unstable.

For site-centered discrete breathers - the ground states of the negative-Hamiltonian pair - we have divided system nonlinearity into two ranges and wrote very good analytic approximations for the breathers in each range.

Lastly, in a dedicated section, we very briefly discuss the specific implementation of the DNLSE in the fields of magnetism, optics, and ultracold atoms, emphasizing ground states. For example, following a 2002 article, we show that the dynamics of a $1d$ optically-trapped ultracold bosonic atoms, in a rather wide range of system densities and system nonlinearities, can be described by a particular version of the here-discussed classical DNLSEs.

## 1. Introduction

A *ground state* is a special state of a physical system. Defined as *the state of lowest energy* [1], the ground state formulation almost always constitutes an essential part in the analysis of physical systems. In [2], for example, the author suggests a search for the ground state as a first step towards understanding of a quantum Hamiltonian. Here we look at systems of coupled discrete nonlinear oscillators arranged in a $1d$ array. The dynamics of these "DNLSE systems" is taken to be governed by the DNLSE. The simplest two-term DNLSE reads [3],[4]:

$$i \cdot \frac{dU_m}{d\zeta} = C \cdot (U_{m-1} + U_{m+1}) + \gamma \cdot |U_m|^2 \cdot U_m$$

(1)

where $\zeta$ is the evolution coordinate (distance or time), $U_m(\zeta)$ is the complex field function of the oscillator at site $m$, the parameter $C$ is the nearest-neighbor coupling constant and $\gamma$ is the unharmonic parameter.

The DNLSE (in numerous variations) appears in the analysis of several physical systems [5]-[13].

DNLSE systems are Hamiltonian systems [14],[15]. The equivalent Hamiltonian from which the two-term DNLSE (Eq. (1)) is derived is made up of two "energy" terms [16]-[18] – a tunneling energy term (designated $\mathcal{H}_2$ below) [19] and an interaction energy term (designated $\mathcal{H}_4$ below) [19].

Depending on the physical system being studied, the tunneling energy term (from which the hopping term in the equation is derived) as well as the interaction energy term (from which the nonlinear term in the equation is derived) may each be either positive or negative. For example, if the physical system studied is an array of light-transmitting optical waveguides close to each other, then the tunneling energy term is positive ($C > 0$) and the interaction energy term is also typically positive ($\gamma > 0$) but can be negative too ($\gamma < 0$), depending on the type of the waveguides' material. If the physical system studied is a gas of trapped interacting ultracold atoms, then the tunneling energy term is negative ($C < 0$) and the interaction energy term is typically positive (repulsive two-body interaction $\Rightarrow \gamma > 0$) but can be negative too (attractive two-body interaction $\Rightarrow \gamma < 0$) ([20], Ch. 22).

In this work we focus our attention on the ground states of these DNLSE systems. We have reviewed existing results sporadically scattered in the literature, provided new insights, and created a comprehensive, coherent picture of the DNLSE ground states. (Somewhat surprisingly, in most of the already-published DNLSE papers, plane wave ground states are not directly or indirectly discussed, and typically not even mentioned. An exception is a 2009 paper considering "minimum energy" and "maximum energy" as part of a comprehensive breathers' stability analysis of a positive tunneling-energy, positive interaction-energy system [21]).

First we classify the four ground state types into two pairs. The instantaneous state of a DNLSE system is described by a set of one-parameter complex functions each with its own amplitude and phase. In consulting the DNLSE Hamiltonians (Eqs. **(5)** and **(6)** below), it becomes apparent that the set of amplitudes minimizing the energy of a positive interaction term (of the Hamiltonian) is very different from the set of amplitudes minimizing the energy of a negative interaction term. Further, we show below that for a given first Hamiltonian and a corresponding first ground state, if a second hamiltonian is created by flipping only the sign of the tunneling energy term, then the corresponding same-energy ground state of the second Hamilton is obtained from the first ground state by flipping the sign of all "odd" amplitudes **[22],[23]**. It is therefore necessary and sufficient to classify the four DNLSE Hamiltonian versions into two pairs: a "positive/negative Hamiltonian pair" *if the sign of the interaction energy term* is positive/negative. The pair of ground states associated with the positive hamiltonian are closely related (odd amplitudes flipped) and have the same energy. Similarly, The pair of ground states associated with the negative hamiltonian are closely related and have the same energy.

Specifically, ground states of the positive Hamiltonian pair are *discrete plane waves* in either a ferromagnetic-like or an antiferromagnetic-like configuration and Ground states of the negative Hamiltonian pair are either unstaggered or staggered *site-centered discrete breathers* **[21]**.

The Hamiltonian-derived DNLSE equation has only two integrals of motion (conserved quantities) **[14],[18],[24]** and therefore, for a system of more than two sites, has no general analytic solution for all possible initial conditions **[16]**. However, for several sets of specific initial conditions, analytic or recursive "specific solutions" do exist **[25]**. One specific solution describes a set of *discrete plane wave states* where the complex field functions are of equal amplitudes and of equally-spaced phases **[18]**, **[24],[26],[27]**. If the phases are all equal (zero phase spacing) **[13],[20]** or are $\pi$ radians apart **[17],[21]** then the two corresponding discrete plane wave states describe the two non-degenerate ground states (disregarding the infinite initial-global-phase degeneracy) of the pair of the positive-Hamiltonian systems.

Another set of specific solutions describes a set of *site-centered discrete breathers* **[21]** (or *single-site peaked discrete soliton (Breather)* **[3]**, or *stationary soliton states* **[28]**, or *nonlinear bound state*, *standing wave* or *stationary state* **[29]**). Mathematically, these breathers are described by recursive equations **[21],[28]**. The breathers may be staggered or unstaggered **[18],[27]**. These two-type breathers describe the two ground state types of the pair of the negative-Hamiltonian systems. Both unstaggered and staggered ground state types are number-of-sites-fold degenerate (disregarding the infinite initial-global-phase degeneracy).

In addition to the Hamiltonian ("energy"), a second conserved quantity of the DNLSE is *density* (*norm*, *number of particles*). A more accurate and complete definition of the DNLSE system's ground state would be – "the smallest value of the Hamiltonian of all possible DNLSE solutions *at a given density*" **[3],[21],[29]**.

We also show below that ground states of the positive-Hamiltonian pair are maximum energy states of the negative-Hamiltonian pair and vice versa.

DNLSE systems are *isolated* classical systems. Once the system is launched with a certain overall density and energy, the system will remain at the launched values of these two quantities at all distances (times). The only change during system evolution is the internal diffusive-redistribution of density and energy among the participating oscillators [22]. These changes are almost always accompanied by a monotonic rise of system's entropy. DNLSE ground states are exceptions. To be in the ground state the system *must be so prepared*. Once the system is in the ground state "nothing happens", except for continuous accumulation of a global phase (quantum-mechanically an unobservable change. We note on passing that unlike DNLSE systems, all quantum mechanical systems undergo quantum fluctuations even at their ground states [30]). Ground states' entropy stays fixed (at a zero value) at all distances (times). As a curiosity, specific-density "frozen" DNLSE states (including specific-density ground states) in which even global phase accumulation does not occur, are presented in the text [25]. Another curious case is a two-site system which, under very specific initial conditions, "glides" into a stationary exited state with a constant-valued entropy during the entire glide [31].

Next we proceed to presenting and discussing some basic properties of the ground states of the positive-Hamiltonian pair - entropy, temperature, correlations and stability. In the stability section we have extended the analysis to include excited plane waves. Our stability analysis of these excited plane waves is based on a specifically engineered perturbation of their phases. Stability is then tested against entropy change. We show that under such specifically engineered phase perturbation, all excited plane waves are entropy-unstable. We note that according to the known linear modulational instability analysis, the lower-energy plane waves are found to be stable [4].

The on-site nonlinearity degree is proportional to the product of the normalized unharmonic parameter and the square of the complex field's amplitude. For the ground states of the negative-Hamiltonian pair - the *s*ite-centered discrete breathers - we take system's nonlinearity to be the nonlinearity associated with the highest-amplitude - the amplitude of the center field (i.e. system's nonlinearity is given as $|\Gamma| \cdot u_o^2$ , see below). We have divided this system's nonlinearity into two ranges and wrote an analytic expression for the breathers in each range. These analytic expressions constitute very good approximations to the exact (recursively calculated) discrete breathers.

Finally, in a dedicated section, we look at the abstract $1d$ DNLSE and its ground state as it appears in the analysis of physical systems in the fields of magnetism, optics and ultracold atoms.

In the following four sections (section **2** to section **5**) we first systematically bring together the key equations and the necessary DNLSE concepts required for the ground states' discussion. Ground states of the positive-Hamiltonian pair and their properties are discussed in the following two sections (section **6** and section **7**). Derivation and approximate analytic expressions for site-centered discrete breathers - the ground states of the negative-Hamiltonian pair - are presented in section **8**. Section **9** is devoted to considerations of the DNLSE in relations to actual physical systems in the fields of magnetism, optics, and ultracold atoms, with eye on the ground states in each case. Our main observations and results are summarized in section **10**.

## 2. Equations

The evolution dynamics of a $1d$ array of (typically) a large number ($N$) of coupled unharmonic oscillators is given by Eq. **(1)** above. Throughout this work, without loss of generality, we will use the notation of optics (evolution coordinate $\zeta$ as distance, or $z$ as a normalized distance, with coupled optical waveguides in mind). The equation consists of two terms – a linear *hopping term*, and a cubic on-site *nonlinear term*. As usual, periodic boundary conditions ($U_{m+N} = U_m$) are assumed here too. In many studies, the linear term includes a diagonal function ($-2 \cdot U_m$) that can be eliminated by a global transformation (moving to a rotating frame) **[3]**.

Several options for normalizing Eq. **(1)** are available **[14]**,**[3]** and are often applied **[12]**,**[13]**. Here, since signs play an important role in our study, we shall eliminate the coupling constant from Eq. **(1)** except for its sign, following a division by $|C|$ –

$$i \cdot \frac{dU_m}{dz} = signC \cdot (U_{m-1} + U_{m+1}) + \Gamma \cdot |U_m|^2 \cdot U_m$$

$$z \equiv |C| \cdot \zeta \; ; \; signC \equiv sign(C) \; ; \; \Gamma \equiv \frac{\gamma}{|C|}$$

**(2)**

In Eq. **(2)**, the evolution coordinate ($z$) – "distance" (or "time") is dimensionless. The normalized unharmonic parameter $\Gamma$, to be referred-to below as a *nonlinear coefficient*, has the units of $[U_m]^{-2}$. If $signC = sign(\Gamma) \; / \; -sign(\Gamma)$ then Eq. **(2)** is a "focusing" / "defocusing" version of the DNLSE **[14]**,**[32]** (also Cf. the optics paragraphs in section **9**).

It is convenient at this point, and indeed done in almost every DNLSE article, to perform a Madelung transformation to the set of density-angle canonical polar variables $(q_m, p_m) \rightarrow (I_m, \phi_m)$ in which the complex field functions $(U_m(z))$ take on the form:

$$U_m = u_m \cdot e^{i \cdot \phi_m} \; ; \; u_m \equiv \sqrt{I_m} \; ; \; \theta_m \equiv \phi_m - \phi_{m+1}$$

**(3)**

Both $u_m(z)$ and $\phi_m(z)$ of Eq. **(3)** are real functions that obey the following dynamics (strictly equivalent to the dynamics of Eq. **(2)**):

$$\frac{du_m}{dz} = signC \cdot u_{m-1} \cdot \{sin(\phi_{m-1} - \phi_m)\} + signC \cdot u_{m+1} \cdot \{sin(\phi_{m+1} - \phi_m)\}$$

$$\frac{d\phi_m}{dz} = -signC \cdot \frac{u_{m-1}}{u_m} \cdot cos(\phi_{m-1} - \phi_m) - signC \cdot \frac{u_{m+1}}{u_m} \cdot cos(\phi_{m+1} - \phi_m) - \Gamma \cdot u_m^2$$

$$u_{m+N} = u_m \; ; \; \phi_{m+N} = \phi_m$$

**(4)**

## 3. Conserved quantities

*Energy*. Equation **(2)** can be derived from a Hamiltonian $(\mathcal{H}_a(\boldsymbol{U}, i \cdot \boldsymbol{U}^*))$ which is a conserved quantity, associated with the system's time translation invariance **[15], [33]** -

$$\mathcal{H}_a(\boldsymbol{U}, i \cdot \boldsymbol{U}^*) = \sum_{m=1}^{N} \left\{ signC \cdot (U_m^* \cdot U_{m+1} + U_m \cdot U_{m+1}^*) + \frac{\Gamma}{2} \cdot |U_m|^4 \right\}$$

**(5)**

The variables $(U_m, i \cdot U_m^*)$ are canonical variables. Adopting the assignment $q_m = U_m$ ; $p_m = i \cdot U_m^*$ **[3],[14],[15],[23],[34]**, equation **(2)** is derived from the Hamiltonian **(5)** as $\frac{dU_m}{dz} = \frac{\partial \mathcal{H}_a}{\partial (i \cdot U_m^*)}$. We note here that if inverted-order assignment of canonical variables is adopted $\left((q_m, p_m) \rightarrow (i \cdot U_m^*, U_m)\right)$ **[17],[35]**, then the sign of the derived DNLSE will be inverted as well. Of course, inverted order of the canonical variables is equivalent to a sign-inverted Hamiltonian **[18]**. We show below that indeed the statistics of DNLSE's is Hamiltonian-sign independent.

However, when it comes to identification of system's ground states, *the sign of the interaction energy term of the Hamiltonian matters*. Note that indeed the sign of the hopping term in **(2)** (and thus the sign of the tunneling energy term in the Hamiltonian) does not play a role in this respect since it can be flipped through the gauge transformation $U_m \rightarrow e^{i \cdot \pi \cdot m} \cdot U_m$ **[22],[23]**. We are therefore making here the important identification of a positive/negative *interaction term* with a "positive/negative-Hamiltonian pair".

In the polar variables of Eq. **(3)** $(I_m, \phi_m)$, the DNLSE Hamiltonian (Eq. **(5)**) takes on the form:

$$\mathcal{H}_2(z) = signC \cdot \sum_{m=1}^{N} 2 \cdot u_m u_{m+1} \cdot \cos\theta_m \quad ; \quad \mathcal{H}_4(z) = \sum_{m=1}^{N} \frac{\Gamma}{2} \cdot u_m^4$$

$$\mathcal{H}_a \equiv \mathcal{H}_2(z) + \mathcal{H}_4(z)$$

$$\hbar_2(z) \equiv \frac{\mathcal{H}_2(z)}{N} \quad ; \quad \hbar_4(z) \equiv \frac{\mathcal{H}_4(z)}{N} \quad ; \quad \hbar_a \equiv \frac{\mathcal{H}_a}{N}$$

**(6)**

In writing **(6)** we have explicitly divided the DNLSE Hamiltonian into its two constituent terms – the nearest-neighbor *tunneling energy* term $(\mathcal{H}_2(z))$ and the on-site *interaction energy* term $(\mathcal{H}_4(z))$. Corresponding site-averaged functions are $(\hbar_2(z), \hbar_4(z))$. Obviously, both $\mathcal{H}_2(z)$ and $\mathcal{H}_4(z)$ vary with propagation distance, but their sum does not. During DNLSE evolution then, an energy diffusion process transfers energy from $\mathcal{H}_2(z)$ to $\mathcal{H}_4(z)$ or the other way around. If the system is prepared in one of its ground states however, diffusions of either density or energy are inhibited and the two Hamiltonian terms stay anchored in their initial values throughout the evolution process.

*Density*. Another conserved quantity of DNLSE systems, thanks to the system's invariance with respect to global phase rotations **[14],[15]**, is "density" ($\mathcal{W}_a$) (or *norm*, or *number of particles*) given by -

$$\mathcal{W}_a = \sum_{m=1}^{N} I_m(z) \;;\; w_a \equiv \frac{\mathcal{W}_a}{N}$$

(7)

The site-averaged values of the two conserved quantities ($w_a, \hbar_a$) form a plane over which a DNLSE phase diagram (not to be confused with "phase space") can be graphically represented.

## 4. Phase diagram, field correlations, system's entropy, system temperature

Next, in relations to ground state properties, we succinctly describe the DNLSE phase diagram and briefly look at several other DNLSE-related quantities.

*DNLSE phase diagram*. The DNLSE phase diagram - **Figure 1** - divides the ($w_a, \hbar_a$) plane into three zones – a lower inaccessible zone, a central thermalization zone, and an upper negative temperature zone **[17]**. The thermalization zone (light purple in **Figure 1**) is bounded from below by the parabola $\hbar_{a,lower-bound}(w_a) = -2 \cdot w_a + \frac{1}{2} \cdot \Gamma \cdot w_a^2$ with minimum of $\hbar_{a,min} = -\frac{2}{\Gamma}$ at $w_{a,min} = \frac{2}{\Gamma}$ and is bounded from above by the parabola $\hbar_{a,upper-bound}(w_a) = \Gamma \cdot w_a^2$. Crossing the thermalization zone is a parabolic "$L_i$ line": $\hbar_{a,L_i\,line}(w_a) = \frac{1}{2} \cdot \Gamma \cdot w_a^2$ **[35]**. Systems initialized with equal amplitudes and uniformly distributed random phases (negligible tunneling energy) fall on the $L_i$ line **[12],[35]**. By definition, ground state systems are placed on the border of the inaccessible zone and the thermalization zone of the DNLSE phase diagram (the blue line of **Figure 1**).

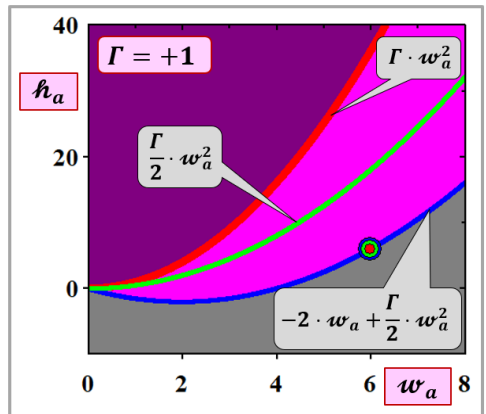

*Figure 1: Phase diagram for a DNLSE system with a positive-Hamiltonian. The colored circular disk on the blue line represents a system with site-averaged density of $w_a = 6$ at its ground state. See the phase diagram of **Figure 10** for a negative-Hamiltonian system.*

*Field correlations*. The DNLSE field correlations are defined as [12] (and see also [13] eq. 3.3) -

$$C_k(z) = \frac{1}{2 \cdot N} \cdot \sum_{m=1}^{N} [U_m^*(z) \cdot U_{m+k}(z) + U_m(z) \cdot U_{m+k}^*(z)]$$

(8)

Or, in polar coordinates -

$$C_k(z) = \frac{1}{N} \cdot \sum_{m=1}^{N} u_m(z) \cdot u_{m+k}(z) \cdot \cos[\theta_{m,k}(z)]$$

$$\theta_{m,k}(z) \equiv \phi_m(z) - \phi_{m+k}(z)$$

(9)

Note that $C_0 = w_a$ and $C_1(z) = signC \cdot \frac{\mathcal{H}_2(z)}{2 \cdot N} = \frac{\hbar_2(z)}{2}$. In ([37], Eq. 28) the authors show that for uncorrelated relative angles $(\theta_k)$, as is the case for the coupled oscillator array analyzed here -

$$\langle C_0 \rangle = \langle I \rangle \ ; \ \langle C_k \rangle = \langle \sqrt{I} \rangle^2 \cdot \langle \cos \theta \rangle^k \quad k \geq 1$$

(10)

The expectation ($\langle \cdot \rangle$) of the field correlations in Eq. (10) is over many realizations, given "statistical excitations". Namely – the initialization of the oscillators in the array is under specific statistical restrictions of their amplitudes and/or phases. Thus, if the nearest-neighbor phase differences are not flat-distributed (such that $\langle \cos \theta \rangle \neq 0$), the fields are correlated and correlations exponentially decay with site separation.

Note that if all amplitudes are nearly equal then $\langle I \rangle \approx \langle \sqrt{I} \rangle^2$ and the normalized field correlations ($\langle C_k \rangle / \langle C_0 \rangle$) is given by the correlation of the relative phases ($\langle \cos \theta \rangle^k$). The relative phase angle $\theta_m$ is the relative orientation of the 2D complex "vectors" $U_m$ and $U_{m+1}$. Thus, for nearly equal amplitudes, the normalized DNLSE field correlations ($\langle C_k \rangle / \langle C_0 \rangle$) are phase correlations much like the two-point correlations of spins in spin arrays [38], or like its analogous one-body density matrix for trapped ultracold atoms [39], or like the phase coherence of a Bose–Einstein condensate in a lattice potential [40].

A field-correlations example for positive-Hamiltonian DNLSE systems is shown in ***Figure 2***.

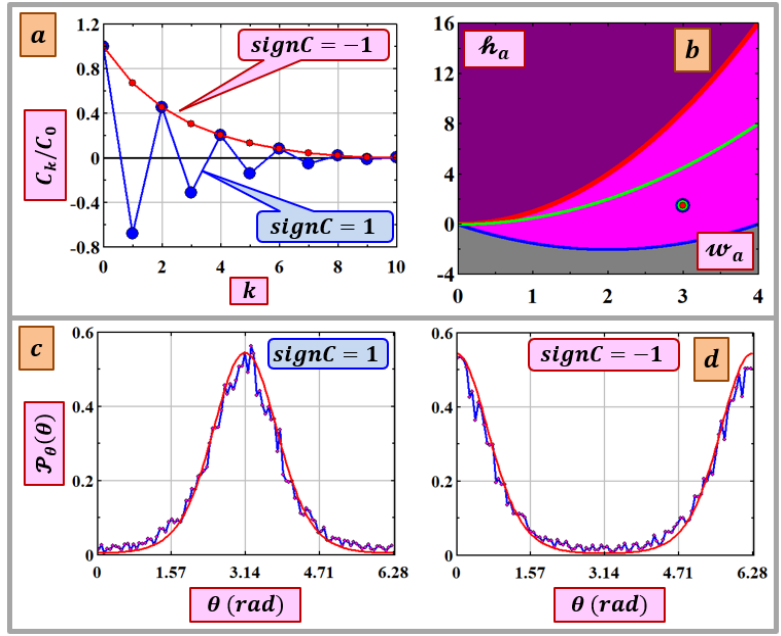

*Figure 2:* *DNLSE field correlations for a positive-Hamiltonian system ($\Gamma = 1$). a. Normalized field correlations vs. site separation (Eqs. (8)-(10), data points obtained by simulation). Field correlations decay exponentially with site separation [12],[37] starting from $k = 1$ (Eq. (10)). b. Position of the system on the phase diagram – half way between the minimum energy line (blue) and the $L_i$ line (green). c,d. PDFs of relative phase angles: π-centered for signC = 1 (focusing) and zero-centered for signC = -1 (defocusing) (compare with the results in [40]. The red curves are theoretical [35].*

*System's entropy.* The coordinates for calculating system's entropy are $(I_j, \theta_k)$ [19]. Following [12],[35],[37], we will adopt here the quantum phase approximation [41] and numerically calculate system's *site-averaged* entropy $(s_{sys})$ as the sum of site-averaged intensity entropy $(s_I)$ and site-averaged entropy of relative phases $(s_\theta)$: $s_{sys} = s_I + s_\theta$.

For numerical calculations, let us divide the range $[0, I_{max}]$ into $M_I$ bins and divide the range $[0, 2\cdot\pi)$ into $M_\theta$ bins. Now define an $M_I \times M_\theta$ array (hundreds by hundreds) and fill the 2d bins with the histograms $\mathcal{D}_{j,k}$ of the number of the corresponding $I_j, \theta_k$ values to get:

$$\mathcal{P}_{j,k} = \frac{\mathcal{D}_{j,k}}{\mathcal{Z}} \ ; \ \mathcal{Z} = \sum_{j=1}^{M_I}\sum_{k=1}^{M_\theta} \mathcal{D}_{j,k} \ ; \ \mathcal{P}_{I,j} = \sum_{k}^{M_\theta} \mathcal{P}_{j,k} \ ; \ \mathcal{P}_{\theta,k} = \sum_{j}^{M_j} \mathcal{P}_{j,k}$$

(11)

Given the probabilities of Eq. (11), Gibbs entropies are calculated as [42] -

$$s_{sys} = -\sum_{j,k} \mathcal{P}_{j,k} \cdot \ln(\mathcal{P}_{j,k}) \ ; \ s_I = -\sum_j \mathcal{P}_{I,j} \cdot \ln(\mathcal{P}_{I,j}) \ ; \ s_\theta = -\sum_k \mathcal{P}_{\theta,k} \cdot \ln(\mathcal{P}_{\theta,k})$$

(12)

System's entropy so calculated is approximate and is weakly dependent on the (arbitrarily selected) number of bins, but the general entropy trends are preserved.

*System Temperature*. DNLSE system temperature was defined in **[35]** as $T_{DNLSE}(w_a, \hbar_a) = \left(\Gamma \frac{\partial s_{sys}(w_a, \hbar_a)}{\partial \hbar_a}\right)_{w_a}^{-1}$. For any system nonlinearity (even at $|\Gamma| \cdot w_a \to 0$) the DNLSE temperature everywhere on the thermalization zone was shown in **[43]** to be given as $T_{DNLSE} = \frac{1}{\beta \cdot \Gamma}$, where $\beta$ is a Lagrange parameter and $sign(\beta) = sign(\Gamma)$. At all system nonlinearities the thermalization zone of the DNLSE phase diagram is bounded from below by a zero temperature line ($\beta \to \infty$) and from above by an infinite temperature line ($\beta \to 0$) **[17]** (blue and red lines of ***Figure 1*** respectively).

*A note on dimensions*. The dimension of the DNLSE Hamiltonian (Eq. **(5)**) is equal to the dimension of the complex site functions squared – say $Q$. The dimension of the nonlinear coefficient ($\Gamma$) is $Q^{-1}$. The dimension of the Lagrange parameter $\beta$ is $Q^{-1}$ as well. The dimension of the DNLSE temperature is then $Q^2$. In short, DNLSE "energy" is measured here in $Q$ and DNLSE temperatures, clearly not the classical thermodynamic temperatures, are measured in $Q^2$.

The dimension $Q$ vary from case to case. In optics, $Q$ stands typically for power/volume **[4],[27]**,[4] (also in Yariv's coupled-mode theory **[44]**). In a coupled pendulums analysis $Q$ is $angle^2$, and in a coupled springs analysis it is $length^2$. In many studies, if the complex field functions ($U_m's$) are dimensionless by definition **[13]**, or are normalized **[12]**, or if the field functions represent probability amplitudes **[3]**-**[8],[22],[45],[46]**, or if the field functions represent discrete order parameters, (**[20]** Ch. 22), **[47]**, then $Q$ is dimensionless.

## 5. Discrete plane wave states

Discrete plane waves states are states of the oscillator array that can be described by analytic solutions of the DNLSE. In each plane wave state (numbered $q$) all complex field functions $\left(U_{m,q}(z)\right)$ are of the same amplitude ($u_0 > 0$) and the relative phases $\left(\theta_{m,q} = \phi_{m,q} - \phi_{m+1,q} \equiv \alpha_q\right)$ are all equal as well **[4],[14],[18],[26],[27],[48]**:

$$U_{m,q}(z) = u_0 \cdot e^{i \cdot m \cdot \alpha_q} \cdot e^{-i \cdot k_{zq} \cdot z}$$

$$\alpha_q \equiv \frac{q}{N} \cdot 2 \cdot \pi \; ; \; q \; integer \; ; \; k_{zq} = signC \cdot 2 \cdot cos\alpha_q + \Gamma \cdot u_0^2$$

**(13)**

The values of the discrete angles $\left(\phi_{m,q}\right)$ are fixed so as to satisfy the periodic boundary conditions. On propagation, all field functions accumulate phase at the same rate $\left(k_{zq} \cdot z\right)$ and thus the *relative* phases stay constant at their initial value. Entropy of all discrete plane wave states is zero. The two ground states of a the positive-Hamiltonian pair are special members of the family of discrete plane wave states (with $\alpha_q = 0$ or $\alpha_q = \pi$).

A small density-preserving perturbation (cannot be applied to the ground states) will cause the discrete plane wave states to thermalize. Below we shall get back to this stability issue in relation to ground state properties.

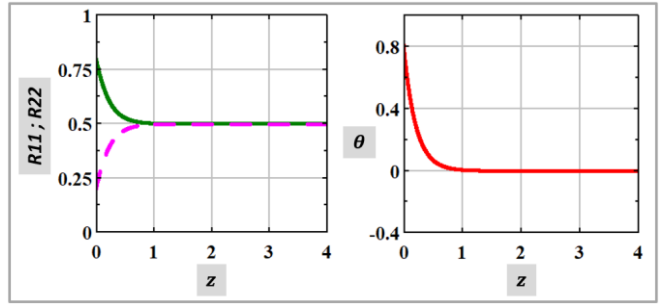

*Figure 3:* The "glide" of a two-site system into a plane wave state. Left – normalized densities. Right – evolution of the relative phase. Under a carefully prepared set of initial conditions, a two-site system with a positive-Hamiltonian and $signC = sign(\Gamma)$ (focusing) will glide into a plane wave state of equal amplitudes in a ferromagnetic-like configuration **[49]** (zero relative angle – not a ground state). During the evolution process, tunneling energy $(\mathcal{H}_2(z))$ monotonically grows while interaction energy $(\mathcal{H}_4(z))$ monotonically shrinks, but system's entropy stays constant (at site-averaged value of $\ln(2)$) **[31]**.

For a DNLSE system to be in a zero-entropy discrete plane wave state, it must be initialized into such state. An interesting exception is a two-site system that "glides", ever so "slowly", towards the high energy discrete plane wave state (not the ground state) following a careful initialization into a non-discrete-plane-wave state (*Figure 3*) **[31]**.

## 6. Ground states of the two positive-Hamiltonian systems

Two of the plane wave states of Eq. **(13)** are the ground states of the DNLSE positive-Hamiltonian pair. The site-averaged energy $(\hbar_{a,q})$ of the $q$'s plane wave state is $\hbar_{a,q} = signC \cdot 2 \cdot w_a \cdot \cos\alpha_q + \frac{1}{2} \cdot \Gamma \cdot w_a^2$ with $w_a = u_0^2$. The ground state energy $(\hbar_{a,gs+})$ of the positive-Hamiltonian is $\hbar_{a,gs+} = -2 \cdot w_a + \frac{1}{2} \cdot \Gamma \cdot w_a^2$. The difference then is

$$\hbar_{a,q} - \hbar_{a,gs+} = 2 \cdot w_a \cdot (1 + signC \cdot \cos\alpha_q)$$

(14)

with zeros at $\alpha_q = \pi$ for $signC = 1$ (focusing) and $\alpha_q = 0$ for $signC = -1$ (defocusing). Back to Eq. **(13)**, the two ground states $(U_{m,gs+})$ of the positive-Hamiltonian pair (both of the same energy) are –

$$U_{m,gs+} = \begin{cases} u_0 \cdot e^{i \cdot (2 - \Gamma \cdot u_0^2) \cdot z} & ; \quad signC = -1 \\ u_0 \cdot e^{i \cdot m \cdot \pi} \cdot e^{i \cdot (2 - \Gamma \cdot u_0^2) \cdot z} & ; \quad signC = 1 \end{cases}$$

(15)

Positive-Hamiltonian ($\Gamma > 0$) ground state configuration is ferromagnetic-like **[49]**, if $signC = -1$ and is antiferromagnetic-like **[49]** if $signC = 1$.

Two notes are on order here:

- For the $\alpha_q = \pi$ case (focusing) the number of sites must be even (in order to satisfy the periodic boundary conditions).

- If $\Gamma \cdot u_0^2 = 2$ then there is no phase accumulation and the oscillators stand still right from the start. Graphically, if the functions $\left(U_{m,gs}(\mathcal{H}^+)\right)$ are plotted on the complex plane then during dynamics execution the constant-length "vectors" revolve CW for $\Gamma \cdot u_0^2 < 2$, revolve CCW for $\Gamma \cdot u_0^2 > 2$ and revolution stops altogether for $\Gamma \cdot u_0^2 = 2$.

The ground states of a positive-Hamiltonian system are shown graphically in ***Figure 4***.

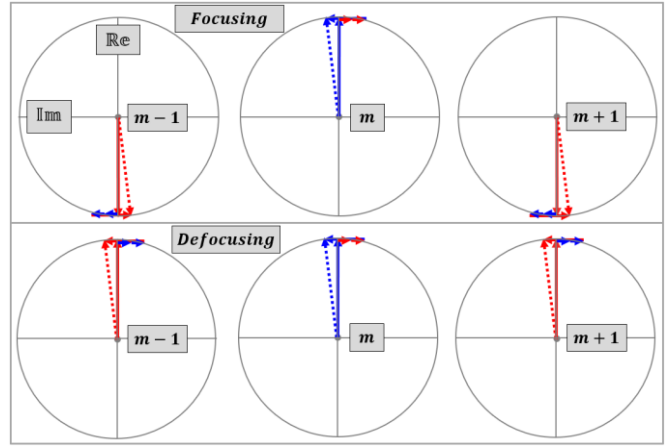

***Figure 4:*** *Ground states of the two positive-Hamiltonian systems (Eq, **(15)**) – graphical illustration. Top: antiferromagnetic-like configuration for a focusing system. Bottom: ferromagnetic-like configuration for a defocusing system. The amplitude of site number $m$ (in the middle) is shown in blue and the amplitudes of its two neighbors are shown in red. Looking at site $m$, the two small red arrows indicate the hopping contribution from the two neighbors to its dynamics, and the single blue arrow indicates the contribution of the on-site nonlinear term to its dynamics. During evolution then, all complex functions of a DNLSE ground state accumulate phase (revolve on the complex plane) at the same rate. Note that amplitude revolution is CCW if the two hopping arrows together are shorter than the single nonlinearity arrow (as shown) or is CW the other way around, or there is no revolution at all in the case of equality.*

The statistical properties of the ground states of the positive-Hamiltonian pair are strictly shared by the corresponding *highest* energy states of the negative-Hamiltonian pair. Evolution of $\hbar_2(z)$ and $\hbar_4(z)$ energies for the two cases is shown in ***Figure 5***. Changing the Hamiltonian sign merely changes the sign of phase accumulation with the evolution distance, and is equivalent to interchanging the position of the two canonical coordinates.

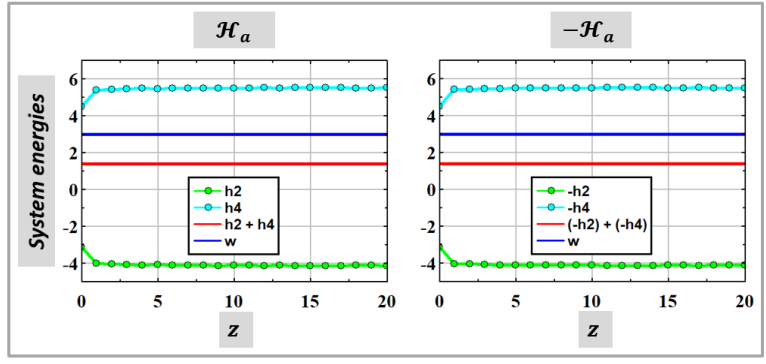

*Figure 5: Effect of changing the Hamiltonian sign. Left: $+\mathcal{H}_a$. right: $-\mathcal{H}_a$. For the curves of the left panel, the system is placed at a certain position on the DNLSE phase diagram. For the curves of the right panel, the system is placed on an equivalent location of the inverted phase diagram (exactly a vertical mirror image of the upright location on the phase diagram (with $h_a = 0$ as the mirror line)). The ground states of the positive-Hamiltonian ($-2 \cdot w_a + \frac{1}{2} \cdot \Gamma \cdot w_a^2$ ; $\Gamma > 0$) become the highest energy states of the negative-Hamiltonian ($= 2 \cdot w_a - \frac{1}{2} \cdot |\Gamma| \cdot w_a^2$ ; $\Gamma < 0$). A ground state and its mirrored highest energy state possess the same statistical properties. In terms of the DNLSE, switching the Hamiltonian sign is equivalent to flipping the position of the canonical variables.*

Let us review some basic properties of the positive-Hamiltonian ground states.

### 7. Basic properties of the ground states of the positive-Hamiltonian pair

Given the expressions for the ground states of the positive-Hamiltonian pair (Eq. **(15)**), we can review some of its basic properties (shared by the highest energy states of the negative-Hamiltonian pair).

*Entropy*. In terms of PDF's, the intensities and angles of each ground state are described by two delta functions: $\mathcal{P}_I(I) = \delta(I - u_0^2)$ and $\mathcal{P}_\theta(\theta) = \delta(\theta)$ or $\mathcal{P}_\theta(\theta) = \delta(\theta - \pi)$. Calculating Gibbs entropy by Eq. **(12)**, we find for the ground states **[21]**, throughout system's evolution -

$$s_{sys}(ground\ states) = 0$$

**(16)**

*Temperature*. Ground state systems of the positive-Hamiltonian pair with energy given by $\hbar_{a,gs+}(w_a) = -2 \cdot w_a + \frac{1}{2} \cdot \Gamma \cdot w_a^2$ are placed on the $\beta = \infty$ line of the DNLSE phase diagram **[17]**. It follows from $T_{DNLSE} = \frac{1}{\beta \cdot \Gamma}$ that

$$T_{DNLSE}(ground\ states) = 0$$

**(17)**

Note that Eq. **(17)** holds for all distances (times) and for all nonzero values of the unharmonic parameter ($\Gamma > 0$).

Another way to realize Eq. **(17)** is through the width of the ground state's $\mathcal{P}_I(I)$. It was shown in **[35]** that for systems on most of the area of the thermalization zone the equilibrium $\mathcal{P}_I(I)$ is of a Gaussian shape and that temperatures of systems after thermalization are given by the variance ($\sigma_I^2$) of the equilibrium $\mathcal{P}_I(I)$. For the ground states, $T_{DNLSE}(ground\ states) = \sigma_I^2(ground\ states) = 0$ everywhere on the thermalization-zone's lower border and at all evolution distances (times).

*Field correlations.* According to Eqs. **(8)** and **(9)** (and see also Eq. **(10)**), system field correlations $\left(C_{k,gs+}(z)\right)$ of the ground states **(15)** at all distances are

$$C_{k,gs+} = \begin{cases} (-1)^k \cdot w_a \ ; \ signC = 1 \ ; \ k = 0,1,2,... \\ w_a \ \ \ \ \ \ \ \ ; \ \ \ \ signC = -1 \ ; \ k = 0,1,2,... \end{cases}$$

**(18)**

Since $C_{0,gs+} = w_a$, we get for the normalized field correlations:

$$\frac{C_{k,gs+}}{C_{0,gs+}} = \begin{cases} (-1)^k \ ; \ signC = 1 \ ; \ k = 0,1,2,... \\ 1 \ \ \ \ \ ; \ signC = -1 \ ; \ k = 0,1,2,... \end{cases}$$

**(19)**

Note that $\hbar_{2,gs+} = -2 \cdot w_a = -2 \cdot |C_{k,gs+}|$.

It follows from Eqs. **(18)** and **(19)** that the field correlations of the ground states of the positive-Hamiltonian pair do not decay and therefore the field-correlation length of these ground states is infinite:

$$\left(\frac{|C_{k,gs+}|}{C_{0,gs+}}\right)_{(N,k)\to\infty} = 1$$

**(20)**

*Stability.* The stability of discrete plane wave states of DNLSE systems is often judged based on a linear modulational instability analysis **[3],[4],[18],[26],[27]** (first experimental observation of discrete modulational instability in any physical system), **[48]**. The analysis, that is of interest in some physical context, predicts modulational instability of discrete plane wave states with relative phase angle $\alpha_q$ as -

$$System\ unstable\ if\ \begin{cases} focusing\ and & 0 \le |\alpha_q| < \frac{\pi}{2} \\ defocusing\ and & \frac{\pi}{2} \le |\alpha_q| \le \pi \end{cases}$$

**(21)**

provided that the system nonlinearity ($|\Gamma| \cdot w_a$) exceeds a certain critical value ($|\Gamma| \cdot w_a >\approx 2$) **[18],[26],[27]**. But in general, one cannot conclude from a linear stability analysis that a solution is fully stable. Stability of a state by such analysis only indicates that small perturbations cannot grow exponentially with distance (time) **[3]** (remain small with propagation distance). It is worth mentioning at this point that in the modulational instability analysis, energy (of the "pump", i.e. of the unperturbed plane wave) is not conserved. The authors of **[50]**, analyzing modulational instability in the evolution of a small perturbation propagating in an optical fiber along with a strong CW signal, introduced a two-region model to enable energy conservation.

Here we shall adopt an entropy-instability criterion. Depending on initial conditions, DNLSE systems may drift into equilibrium in what seems to be an entropy-driven process. If the system is initialized into the thermalization zone of the phase diagram, then the system will drift into a thermal equilibrium **[17],[35]**. During the drift to equilibrium, both $\mathcal{P}_I(I)$ and $\mathcal{P}_\theta(\theta)$ continuously change their shapes (and hence the values of the energies $\hbar_2(z)$ and $\hbar_4(z)$ continuously change their values as well). The panels of ***Figure 6*** show an example of $\mathcal{P}_I(I)$ and $\mathcal{P}_\theta(\theta)$ evolutions for a DNLSE on the thermalization zone. These evolutions of $\mathcal{P}_I(I)$ and $\mathcal{P}_\theta(\theta)$ are (almost always) accompanied by a monotonic rise of system's entropy from an initial (low) value to a higher equilibrium value.

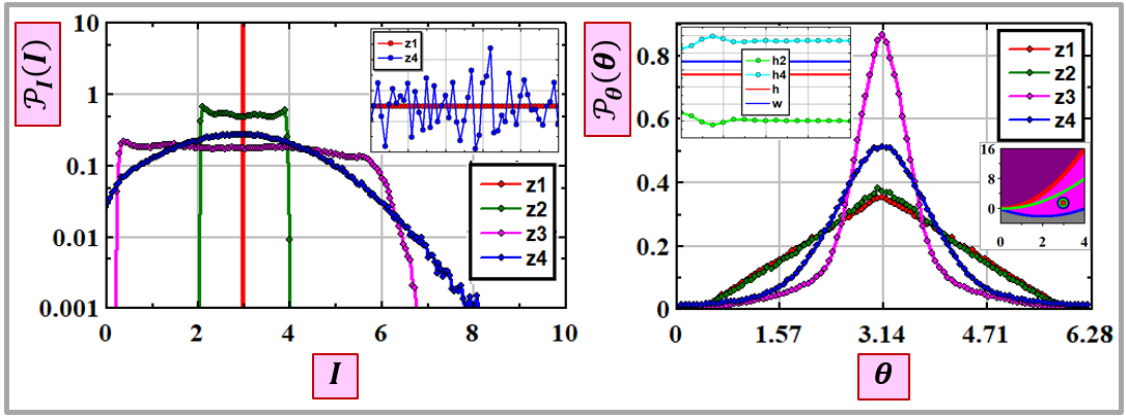

***Figure 6:*** *Evolution of PDFs for DNLSE systems on the thermalization zone (position shown by the phase diagram inset on the right). The panels show four snapshots taken at $z1 = 0$ to $z4 = z_{end}$ of $\mathcal{P}_I(I)$ (left) and of $\mathcal{P}_\theta(\theta)$ (right). The inset on the left panel shows intensities at $z1$ and $z_{end}$. The green and blue curves of the inset on the right show the evolutions of $\hbar_2(z)$ and $\hbar_4(z)$ respectively. As expected for these PDFs (verified numerically but not shown in the figure), at short distances the relative-angle entropy ($s_\theta$) goes down, the density entropy ($s_I$) shoots up, and the system's entropy (the sum of the two) monotonically rises. Note the overshoot of $\mathcal{P}_\theta(\theta)$ at $z3$. In this work we have adopted entropy-change as a criterion for stability of DNLSE systems.*

Mathematically, discrete plane wave states (Eq. **(13)**) evolving with delta functions $\mathcal{P}_I(I)$ and $\mathcal{P}_\theta(\theta)$, independent of distance, seem to present an exception (to thermalization). However, If a small perturbation (either phase noise or amplitude noise) is introduced to these plane wave states, the initiated delicately balanced arrangement of the site-fields will be interrupted and the now slightly wider delta probabilities will grow wider with distance until their equilibrium shapes are reached.

For example, a pronounced instability as indicated by the spread of field amplitudes (increased width of the $\mathcal{P}_I(I)$ function) upon a small phase perturbation is shown for two discrete plane wave states by the panels of ***Figure 7***. And wider PDFs mean higher entropies. We shall refer to this kind of instability of DNLSE states as *entropy-instability*.

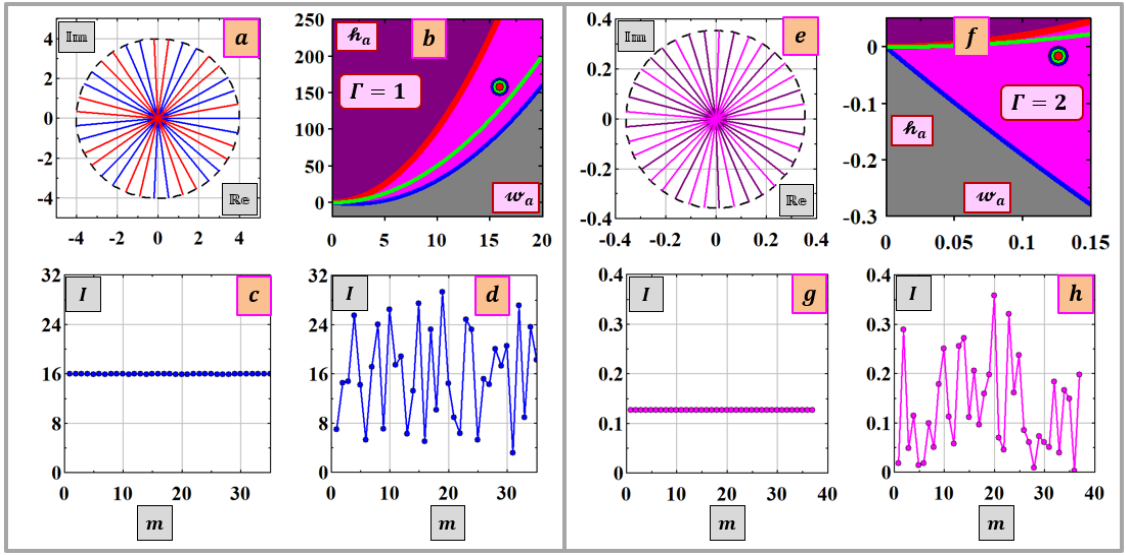

*Figure 7: Entropy-instability of two DNLSE discrete plane wave states. a. The equally-spaced amplitudes on the complex plane ($\alpha_q = 0.36$). b. Rather high location of system a on the phase diagram. c. Site-densities ($I_m$) at $z = 0$. Random phase noise added, flat-distributed between $\pm 0.032$. d. Wide-spread site-densities at a distance ($z = 0.6$), indicating system's entropy-instability. e. The equally-spaced amplitudes on the complex plane ($\alpha_q = 1.7$). The amplitudes $u_0$ are rather small ($u_0^2 = 0.127$), and the plane wave state is "frozen" (no phase accumulation) since the value of the amplitudes was purposely so selected (to satisfy $\Gamma \cdot u_0^2 = -signC \cdot 2 \cdot \cos \alpha_q$, Eq. (13)). f. The state is located low, below the $L_i$ line (green in the figure) on the phase diagram. g. Site-densities ($I_m$) at $z = 0$. Random phase noise added, flat-distributed between $\pm 0.16$. h. Wide-spread site-densities at a distance ($z \cong 250$), indicating system's entropy-instability.*

Next, let us take a close look at entropy changes of discrete plane wave states under a carefully designed "location-preserving" perturbation. We propose to engineer a perturbation that will preserve *both* conserved quantities of a "q" plane wave state and thus preserve the location of the state on the DNLSE phase diagram (same location before and after the perturbation). In other words – the engineered location-preserving perturbation will not only preserve the site-averaged density of the state ($w_a$) but will also preserve its site-averaged energy ($\hbar_a$).

To this end, it is immediately clear that a phase-only perturbation will preserve both the site-averaged density ($w_a = u_0^2$) of the state and its site-averaged interaction energy $\left(\hbar_4 = \frac{1}{2} \cdot \Gamma \cdot u_0^4\right)$. To preserve the state's site-averaged tunneling energy $\left(\hbar_2 = signC \cdot 2 \cdot u_0^2 \cdot \cos \alpha_q\right)$, a set of random relative angles $\theta_{m,rand}$ is generated within a narrow range near $\alpha_q$ with an imposed constraint: $\langle \cos \theta_{m,rand} \rangle = \cos \alpha_q$. The result then is the sought-for location-preserving perturbation.

The proposed instability criterion of a discrete plane wave state under such location-preserving perturbation is the change in system's entropy during evolution. A rise in state's entropy indicates state's instability. In *Figure 8* we show the entropy change of two discrete plane wave states under location-preserving phase perturbations. As shown, both plane wave states are unstable according to the entropy-instability criterion, even though one of the states is predicted to be stable according to the linear modulational instability analysis.

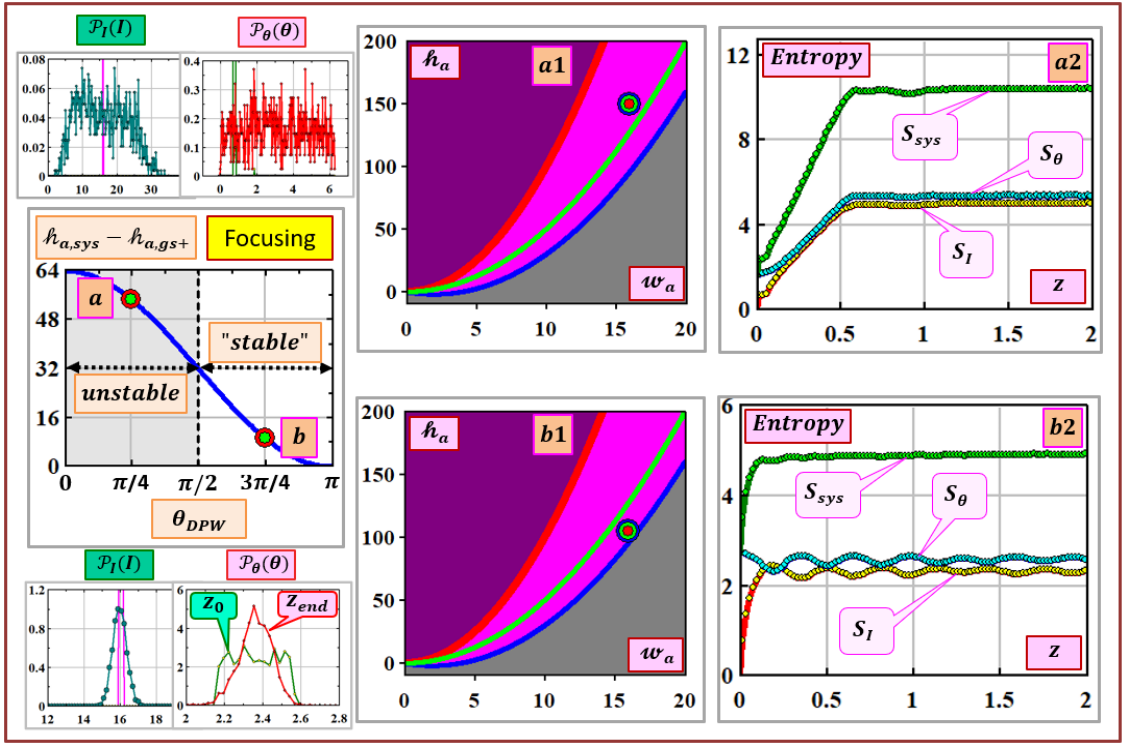

*Figure 8:* *Entropy-instability of discrete plane wave states. Two states of a focusing DNLSE are considered – state a $\left(u_o = 4\,;\alpha_q = \frac{\pi}{4}\right)$ and state b $\left(u_o = 4\,;\alpha_q = \frac{3\cdot\pi}{4}\right)$, as shown on the center panel on the left. State a is high on the thermalization zone of the DNLSE phase diagram (panel a1) and state b is rather close to the lowest energy level (panel b1). According to the linear modulational instability analysis, plane wave b is stable (Eq.* **(21)***). Both states are perturbed by a weak location-preserving phase perturbation (see text) with $\mathrm{stdev}(\cos\theta_{m,rand})/\mathrm{mean}(\cos\theta_{m,rand}) = 0.044$. Evolution snapshots of $\mathcal{P}_I(I)$ and $\mathcal{P}_\theta(\theta)$ are shown by the panels on the left (top / bottom for plane wave a / b) and continuous entropy evolutions are shown by panels a2 and b2 on the right. The figure shows that not only discrete plane wave a is entropy-unstable but even discrete plane wave b is entropy-unstable as well.*

The entropy-instability test under a location-preserving perturbation can be applied to each and every plane wave state of a finite length system (finite $N$) of the positive DNLSE Hamiltonian pair, *except for the two ground states*. Since the ground states are at extrema of the cosine function ($\alpha_q = \pi$ for a focusing system and $\alpha_q = 0$ for a defocusing system), a location-preserving perturbation does not exist. In other words, *any* perturbation to a ground state of a positive-Hamiltonian DNLSE system will excite the state to a higher energy level.

To conclude the stability discussion – except for the two ground states, all other (excited) plane wave states of a finite system of the positive DNLSE Hamiltonian pair, are entropy-unstable under a location-preserving perturbation. The ground states cannot be perturbed this way and thus considered stable.

## 8. Ground states of the two negative-Hamiltonian systems

The ground states of an infinite ($N = \infty$) negative-Hamiltonian system ($\Gamma < 0$), are known as a *site-centered discrete breathers* [21]. (For other common names see the introduction above or see [3],[28],[29],[48]). A site-centered discrete breather is a symmetric field structure with a maximum amplitude centered on a specific site along with fast decaying amplitudes on both of its sides. Practically therefore, these breathers represent the ground states of "correctly truncated" finite systems (*Figure 9*). These ground state breathers, all of the same energy for a given density, could be unstaggered or could be staggered, depending on the value of the $signC$ parameter of the DNLSE (focusing / defocusing – cf. *Figure 9*). Both breather types are *stationary solutions* of the DNLSE [16]. The ground states of a negative-Hamiltonian DNLSE system, either focusing or defocusing, are $N$-fold degenerate since each and every site of the $N$-long oscillator array can be the center of a ground state breather.

As discussed above in relations to ground states of the positive-Hamiltonian pair, site-centered discrete breathers are also the *maximum* energy states for an infinite *positive-Hamiltonian* system ($\Gamma > 0$) [21]. Note that for a fixed value of the site-averaged density ($w_a$), the site-averaged energy ($\hbar_a$) goes to infinity with the length of the array $\left(w_a = const. \Longrightarrow \hbar_a(N) \xrightarrow{N \to \infty} \infty\right)$. As site-averaged density grows, the single site centered breather gets increasingly concentrated about one lattice site [29],[47].

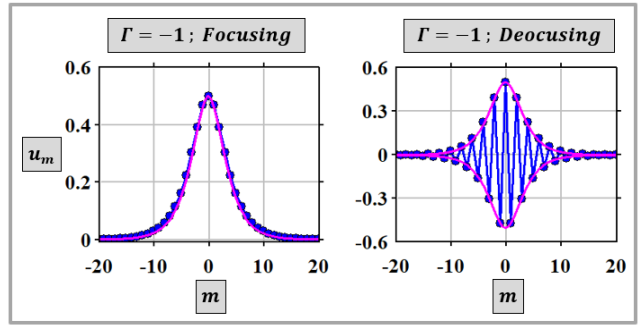

*Figure 9: Ground states of negative-Hamiltonian systems. Shown in the figure are two site-centered discrete breathers – unstaggered (left) and staggered (right). These breathers are "stationary solutions" of the DNLSE. The shown breathers are relatively shallow and thus relatively wide (extend to more than ten sites to each side of the maximum-amplitude site). The blue dots were calculated through the recurrence relations of Eq. (22) below. The red lines are matched hyperbolic secant envelopes. Actual values of the breather's amplitudes are well approximated by the hyperbolic secant expression of Eq. (23) below.*

All fields ($U_m's$) of a discrete breather uniformly oscillate, i.e. they accumulate their phase at the same rate. All field amplitudes of a discrete breather ($u_m's$) are independent of the evolution distance. The fields of a discrete breather are given then by $U_{m,gs-}(z) = u_m(0) \cdot e^{-\kappa \cdot z}$. With initial relative phases of $\theta = 0$ ($\theta = \pi$), the tunneling energy of the system ($\mathcal{H}_2(z)$, Eq. (6)) is minimized for $signC = -1$ ($signC = 1$). The launched amplitudes ($u_m$'s) must then satisfy the derived recurrence relations (22) below (cf. Eq. (4)), with an added restriction of $u_m \xrightarrow{|m| \to \infty} 0$ [48]:

$$u_{m+2} = signC \cdot \kappa \cdot u_{m+1} - u_m - signC \cdot \Gamma \cdot u_{m+1}^3 \ ; \ m = 0,1,2,...$$

$$\kappa = \frac{signC \cdot 2 \cdot u_1 + \Gamma \cdot u_0^3}{u_0} \ ; \ u_m \xrightarrow{|m| \to \infty} 0$$

$$u_{-m} = u_m \ ; \ u_0 > 0 \ ; \ u_1 > 0 \ if \ signC = -1 \ ; \ u_1 < 0 \ if \ signC = 1$$

(22)

According to Eq. (22), once the maximum amplitude $u_0$ (at $m = 0$) is known, the two neighboring amplitudes ($u_1 = u_{-1}$) are determined through the decay to zero requirement, and then $u_m$'s for $|m| > 1$ are determined through the recurrence relations in (22). In [16], similar recurrence relations are presented in a matrix form. In [28], similar recurrence relations were stated as localized solutions to the DNLSE of the form $u_m(\kappa) \cdot e^{-i \cdot \kappa \cdot z}$, including the decay to zero at infinity requirement. In [21], studying a positive-Hamiltonian system, recurrence relations similar to recurrence relations (22) (for extremum energy) were derived through a specific variation of an expression that includes the two DNLSE conserved quantities.

It is worth mentioning here that $u_{m+1} = u_m = u_0$ (ferromagnetic-like) or $u_{m+1} = -u_m$ ; $|u_m| = u_0$ (antiferromagnetic-like) with $\kappa = -2 + \Gamma \cdot u_0^2$ are two possible stationary solutions of Eq. (22), which are the two ground states of the *positive-Hamiltonian* pair systems (cf. Eq. (15)). Of course, in these cases the decay to zero requirement is waved.

Analytic expressions exist, offering very good approximations to the negative-Hamiltonian ground states, depending on how strong system nonlinearity ($|\Gamma| \cdot u_0^2$) is. In the low nonlinearity range ($|\Gamma| \cdot u_0^2 < 1$) wide breather ground states are closely approximated by a hyperbolic secant function. The hyperbolic secant function is well known in describing the fundamental soliton supported by the *continuous* NLSE [27], [48]. In the high nonlinearity range ($|\Gamma| \cdot u_0^2 \gg 1$) narrow breather ground states are closely approximated by a decaying exponential function [21] -

$$U_{m,gs-}(z) = u_m \cdot e^{-i \cdot \kappa \cdot z}$$

$$u_m = \begin{cases} u_0 \cdot (signCG)^m \cdot \text{sech}\left(\frac{m}{\sigma_L(u_0)}\right) & ; \quad |\Gamma| \cdot u_0^2 < 1 \ ; \ m = 0, \pm 1, ... \\ u_0 \cdot (signCG)^m \cdot e^{-\frac{|m|}{\sigma_H}} \ ; \ \sigma_H = \frac{1}{\ln(|\Gamma| \cdot u_0^2)} \ ; \ |\Gamma| \cdot u_0^2 \gg 1 \ ; \ m = 0, \pm 1, ... \end{cases}$$

(23)

The width parameter ($\sigma_L(u_0)$) of the hyperbolic secant in Eq. (23) can be determined by solving an implicit equation based on the recurrence relations (22) for $u_2$ with $signC = -1$.

The hyperbolic secant wide site-centered discrete breathers, unstaggered and staggered ($signCG = signC \cdot sign(\Gamma) = -1$) are shown by the panels of *Figure 9* above. A narrow, staggered, site-centered discrete breather is shown in *Figure 10*. The figure also shows the ground state's location on the DNLSE phase diagram for a finite, arbitrarily selected number of coupled oscillators.

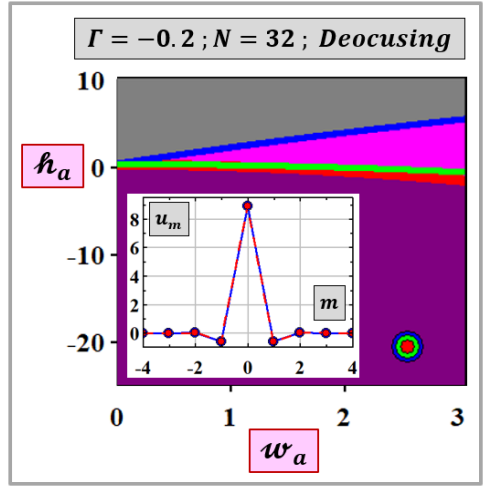

*Figure 10: Ground state of a negative-Hamiltonian system. Shown in the inset is a narrow, staggered, site-centered discrete breather (blue, Eq. (22)), well approximated by a decaying exponential function (red dots given by Eq. (23)). The figure shows the location of the system on the DNLSE phase diagram, deep down in the negative temperature zone, for an arbitrarily selected number of sites in the array ($N = 32$ in the figure). For a fixed site-averaged density ($w_a = 2.52$ in the figure) system's location will drift further down to negative infinity as the number of sites will get increased to infinity.*

The entropy of a negative-Hamiltonian DNLSE system at its ground state is greater than zero, its temperature is negative, it has a finite field-correlation length, and the state is stable [21],[3]. Further analysis of the properties of discrete breather ground states is beyond the scope of the present work.

## 9. Relations to actual physical systems

So far we have considered an abstract DNLSE. In this section we will relate the abstract DNLSE to actual physical systems in the fields of magnetism, optics, and ultracold atoms.

*Magnetism*. Clear similarities can be identified between Hamiltonians of magnetic exchange interactions and the DNLSE Hamiltonian. The Hamiltonian of the classical *XY* model for example, a reduced version of the classical Heisenberg model [49], is practically identical to the tunneling energy term of the DNLSE Hamiltonian. The configuration energy ($\mathcal{H}_{XY}$) of spins in the *XY* model is $\mathcal{H}_{XY} = -2 \cdot J \cdot \sum_i \mathbf{S}_i \circ \mathbf{S}_{i+1} = -2 \cdot J \cdot \sum_i \cos(\theta_i)$ where $\theta_i \equiv \phi_i - \phi_{i+1}$, $\phi_i$ is the angle of orientation of spin vector $\mathbf{S}_i$, constrained to lie in a plane, and the parameter *J* is an exchange interaction (coupling) constant [51]. Here we have taken the spins to be on the sites of a $1d$ lattice and for the *XY* model periodic boundary conditions are typically assumed.

In a more general *XY* model, $J$ depends on the site number ($J \to J_i$) in which case the Hamiltonian ($\mathcal{H}_{XY}$) of the *XY* model is identical to the tunneling energy term of the DNLSE Hamiltonian ($\mathcal{H}_2$) with $signC = -1$ and $J_i = 2 \cdot u_i \cdot u_{i+1} \geq 0$ (Eq. **(6)**).

Given a discrete plane wave state of the positive DNLSE Hamiltonian, the tunneling energy term $\hbar_2 = signC \cdot 2 \cdot u_0^2 \cdot \cos \alpha_q$ is analogous to the site-averaged *XY* Hamiltonian: $\frac{\mathcal{H}_{XY}}{N} = -2 \cdot J \cdot \cos\theta$ with $J_i = J > 0$. If $signC = -1$ then the DNLSE ground state is ferromagnetic-like $(\alpha_q = \theta = 0)$ **[52]**. But the exchange interaction constant could also be negative ($J < 0$) **[49]** corresponding to $signC = 1$. The ground state tunneling energy of the positive DNLSE Hamiltonian would in this case be antiferromagnetic-like $(\alpha_q = \theta = \pi)$. In **[2]**, the spin-spin interaction Hamiltonian is written with a positive sign of the interaction constant $(+J\,; J > 0)$ and therefore the Hamiltonian in **[2]** corresponds to an antiferromagnetic exchange. In **[53]** the authors introduce a flexibility-enabling sign parameter to the nearest-neighbor interaction term, much like the $signC$ parameter of Eq. **(2)** above.

It is worth pointing again here that the gauge transformation $U_m \to e^{i\cdot\pi\cdot m} \cdot U_m$ **[22]**, **[23]** will flip the sign of the tunneling energy term and at the same time will change $\theta$ by $\pi$ (from 0 to $\pi$ or from $\pi$ to 0). Thus, the two ground states of the positive DNLSE Hamiltonian pair depend on $signC$ in a trivial, straight-forward way.

The tunneling energy of the uniform phase $(\theta_m = 0)$ ground state of the positive DNLSE Hamiltonian corresponds to the Hamiltonian of a $1d$ Ising model in magnetism. The long range order of a defect-free one dimensional lattice of Ising spins at zero temperature **[49]** then corresponds to the infinite field-correlation length of the uniform-phase ground state of the positive Hamiltonian system.

The analogy of magnetic systems and DNLSE systems goes even further. Discussing a one-band two term Hubbard model, the author of **[2]** points to the interaction term as giving rise to highly correlated ground states. In the DNLSE systems, the minimization of the interaction energy of a positive-Hamiltonian system drives the equalization of the field amplitudes which, in turn, gives rise to long range (large site-spacings $k$) field correlations.

Actually, the Hamiltonian of the quantum mechanical Bose-Hubbard model, an important theoretical model in analyzing strongly interacting electrons in magnetic materials **[2]** and of interacting bosons on a lattice **[54]** is, *in its classical counterpart*, the DNLSE Hamiltonian of Eq. **(5)** **[47]** (and see also **[13]**).

*Optics*. The Hamiltonian-derived DNLSE (Eq. **(1)**) is introduced in a number of optics studies, theoretical and experimental, in order to analyze electromagnetic-waves propagation in periodic transparent optical structures (**[55]** and references therein). Typically the "optical structure" is an array of single-mode optical waveguides placed or in-glass "laser-carved" close to each other. During light propagation through the array, the evanescent field of one waveguide excites the guided mode of its neighbors. The linear hopping term of the DNLSE, quantitatively describing these excitations, is derived through the well-known coupled-mode theory (CMT) **[44]**, with strong similarity to the single-band tight-binding approximation **[55]**. The effect of on-site (same waveguide) nonlinear propagation of electromagnetic-waves is taken care of by the nonlinear term of the DNLSE.

It is worth mentioning at this point, that Hamiltonian-derived equations are rare in classical optics. Typically, in the absence of currents and in the absence of free charges in the medium and with the adoption of a scalar approximation, the vectorial Maxwell's equations are reduced to a single scalar homogeneous Helmholtz equation **[56]**. The scalar Helmholtz equation is a wave equation from which optical modes of free space are computed **[57]**,**[58]**, and for *periodic* optical structures Bloch functions **[59]** along with optical bands $k_z(k_x)$ **[56]**,**[60]**,**[61]** are derived. The scalar Helmholtz equation then is commonly derived, under certain assumptions, from Maxwell's equations although the authors of **[62]** suggested its derivation from an exact Hamiltonian ray-tracing system. Even the equation known as the "optical Schrödinger equation" **[63]** is a slowly-varying envelope approximation of the scalar Helmholtz equation, and has no Hamiltonian origin. In optics, it turns out, the Hamiltonian-derived DNLSE discussed here is somewhat of an exception.

Back to the DNLSE (Eq. **(1)**), according to the coupled-mode theory, the coupling coefficient $(C)$ is positive with analytic expressions for nearest-neighbor **[64]** and even next-nearest-neighbor coupling **[56]**. The cubic on-site nonlinear term, preceded by the unharmonic parameter $(\gamma)$, is typically associated with the optical Kerr effect. In that case the unharmonic parameter is positive $(\gamma > 0)$ **[65]**, the nonlinearity is referred to as "focusing nonlinearity" (leading to the light self-focusing effect) **[65]**, and the associated interaction energy term of the Hamiltonian is positive. It is important not to confuse "focusing/defocusing" when referring to the version of the DNLSE as a whole with "focusing/defocusing" when referring to the nonlinearity term itself. For example, the author of **[14]**, refers to a *negative* nonlinearity term of the DNLSE (attractive atom-atom interaction in Bose-Einstein condensate (BEC)) as a *focusing* nonlinearity term. In optics, negative nonlinearity would be "defocusing" **[18]**.

The linear part of the DNLSE, derived in optics through the CMT, is a simplification of the more general periodic structure cases. In the more general case, still $1d$, the transverse coordinate of the periodic structure (say $x$) is explicitly included in the dynamic equation, leading to light propagation analysis in terms of two-variable complex functions (say $\psi(x,z)$) known as Floquet–Bloch (FB) functions. In a number FB studies, positive as well as negative nonlinearities are considered or experimented with **[18]**. For example studies with photorefractive materials **[66]**-**[70]**, or with nematic liquid crystals **[71]**,**[72]**, or with Lithium niobate waveguide arrays **[73]**, or with a sodium-vapor **[74]**.

In optics, the CMT-derived hopping term of the DNLSE (Eq. **(1)**) "generates" a single optical band $\left(k_z(\alpha_q) \ ; \ -\pi \leq \alpha_q \leq \pi\right)$ **[75]**. The CMT optical band has a cosine shape, very similar to (but not necessarily identical with) the shape of the first optical band generated through the FB theory for a $1d$ periodic optical structure. See in **[56]**, for example, the comparison of the CMT band with the first FB band calculated for a $1d$ array of step-index waveguides (the Kronig Penney model). To be consistent with the shape of the lower band as calculated by the tight binding model for crystals or for a $1d$ chain of spins **[49]**, the DNLSE for optical waveguides should be written with a negative hopping term $\left\{i \cdot \frac{dU_m}{d\zeta} = -C \cdot (U_{m-1} + U_{m+1}) + \gamma \cdot |U_m|^2 \cdot U_m \ ; \ C, \gamma > 0\right\}$. The equation is derived from a positive-Hamiltonian with $q_j = U_m \ ; p_j = i \cdot U_m^*$ as the canonical variables. The "optical DNLSE" here is of a defocusing type with a focusing optical nonlinearity term. The CMT optical band in this case is $k_z \propto -\cos \alpha_q \ ; \ -\pi \leq \alpha_q \leq \pi$ and the positive-Hamiltonian ground state, at the lowest point of the band $(\alpha_q = 0)$, is ferromagnetic-like. In strict similarity with the ground state of an ultracold bosonic system, as discussed next.

*Ultracold atoms*. In relations to the DNLSE, the most relevant system of ultracold atoms is a system of a large number of bosons interacting with repulsive forces at a low temperature. The bosons are confined by a smooth external potential and trapped in a $1d$ optical lattice potential. The dynamics of such systems is well described by the single-band quantum Bose-Hubbard model [76]. Following [13], let us specify the conditions in which the quantum operators of the Bose-Hubbard model can be replaced by classical complex functions.

Assuming a smooth parabolic external potential, a decisive quantum-to-classical switch parameter is the mean number of bosons $N_B$ in the central site. If the external parabolic potential is very shallow, then the mean number of bosons per site is approximately equal to the *dimensionless* site-averaged density $w_a$, i.e. $N_B \cong w_a$ with $w_a$ as defined in Eq. (6). In terms of the DNLSE parameters $(\Gamma, w_a)$ the quantum-to-classical switch is allowed in a wide range given by $1 \ll \lambda \equiv \Gamma \cdot w_a \ll w_a^2$ [13].

Thus, at large enough system nonlinearity $(\Gamma \cdot w_a)$, but not too large, the low-temperature dynamics of the system can be described by treating the quantum operators as classical complex functions. In the cited conditions then, an adequate description of system's dynamics would be the discrete semiclassical version of the familiar Gross-Pitaevskii (GP) equations for complex functions [13]. The discrete GP Hamiltonian is given as $\mathcal{H}_{GP}(\boldsymbol{U}, i \cdot \boldsymbol{U}^*) = \sum_{m=1}^{N} \left\{ -(U_m^* \cdot U_{m+1} + U_m \cdot U_{m+1}^*) + \frac{V_m}{J} \cdot |U_m|^2 + \frac{\lambda}{2} \cdot |U_m|^4 \right\}$. The parameter $V_m$ preceding the quadratic diagonal term is the value of the external potential at site $m$, and the parameter $J$ is the tunneling amplitude between neighboring lattice sites. The site functions $U_m(z)$ are normalized by $w_a$ (so that their mean value is about unity) and the evolution parameter $(z)$ is in this ultracold-bosons case, a dimensionless time. The discrete GP Hamiltonian is seen to be the DNLSE (Eq. (5)) with an added diagonal confining-potential term. If the confining potential is very shallow or there is no external potential at all, then *the dynamics of a $1d$ optically-trapped ultracold bosonic atoms, in a rather wide range of system densities and system nonlinearities, can be described by the classical DNLSE*.

The upper limit of the quantum-to-classical switch validity $(\Gamma \cdot w_a = w_a^2)$ is a strong-interaction limit above which the ground state of the system undergoes a quantum phase transition from a superfluid to a Mott insulator [13]. Above this upper limit then, the description of the system's dynamics by the discrete GP equations is invalid.

Without the external confining potential, the discrete GP Hamiltonian becomes equal the DNLSE Hamiltonian of Eq. (5). The sign of the hopping term in the Hamiltonian is negative (as is always the case with atomic systems when the contribution of the off diagonal-functions is given by a *negative* discrete second-order derivative), and the interaction energy term is positive for the assumed repulsive atom-atom interaction. Thus, the derived DNLSE for the optically trapped ultracold bosonic system is of a defocusing type with a focusing nonlinearity term (in optics terminology). The ground state of this positive-Hamiltonian ultracold bosonic system is ferromagnetic-like.

Insights to the more general analysis of ultracold atoms *in optical lattices*, classical or quantum [77], can still be gained by considering the simpler DNLSE analysis. An interesting and a rather relevant example is the generation and observation of a superfluid with *tunable ground states* (in momentum space) [78]. The authors of [78] studied and experimented with smoothly confined ultracold atoms in a deep optical lattice. Their employed key experimental technique was a single-parameter-controlled frequency modulation of one of the lattice beams. The driven system can be described to a good approximation by an effective time-independent Hamiltonian that reads, in $1d$ and nearest-neighbor interaction -

$\widehat{\mathcal{H}}_{eff} = -2 \cdot \sum_m |J^{eff}_{m,m+1}| (e^{i\cdot\theta_m} \cdot \hat{a}^\dagger_m \hat{a}_{m+1} + e^{-i\cdot\theta_m} \cdot \hat{a}^\dagger_{m+1} \hat{a}_m) + \widehat{\mathcal{H}}_{on-site}$ with $\theta_m \equiv \phi_m - \phi_{m+1}$ and $\hat{a}^\dagger_m, \hat{a}_m$ denote the creation and annihilation operators. The parameter $|J^{eff}_{m,m+1}|$ is an effective amplitude of tunneling between neighboring sites $m$ and $m+1$ (in either directions). The eigenstates of this effective Hamiltonian are Bloch waves with the dispersion relation $E(k) = -2 \cdot |J^{eff}| \cdot \cos((k-k')\cdot d)$ ([78] with a small modification). The parameter $d$ is the lattice spacing. The parameter $k'$ is a "quasi-momentum" given by $k' = A/\hbar$ where $A$ is an apparent gauge vector potential due to the frequency-modulation of the lattice beam. The lowest energy band then is of a negative cosine shape with a shifted minimum to $k = k'$. The frequency modulation technique thus allows for the generation of superfluid ground-states at a finite and *tunable* quasi-momentum.

## 10. Summary

The DNLSE systems of coupled nonlinear oscillators considered here are classical Hamiltonian systems. The Hamiltonian of the simplest DNLSE system is the sum of two terms – a the nearest-neighbor tunneling energy term and an on-site interaction energy term (Eqs. **(5)** and **(6)**). As we have determined in this work, the four possible Hamiltonian versions must be grouped into two pairs *according to the sign of the on-site interaction energy term*. We thus refer to the pair of DNLSE Hamiltonians composed of a positive/negative interaction energy term as "positive/negative-Hamiltonian pair".

The DNLSE Hamiltonian is described by a set of single-variable complex field functions $(U_m(z))$, each associated with a single oscillator site. The equations of motion are derived from the Hamiltonian through the canonical set $(q_m, p_m) \to (U_m, i \cdot U^*_m)$ or equivalently by the canonical polar variables $(q_m, p_m) \to (I_m, \phi_m)$.

Given the $(U_m, i \cdot U^*_m)$ coordinates, a two-term dynamic equation is derived – a linear hopping term, and a cubic on-site nonlinear term (Eqs. **(1)** and **(2)**). Importantly, the sign of the hopping term can be flipped through the gauge transformation $U_m \to e^{i\cdot\pi\cdot m} \cdot U_m$ [22],[23]. It follows that the ground states for the positive/negative Hamiltonians come in pairs – one ground state of a given pair is obtained from the other by the sign-flipping transformation. The relative angle $(\theta)$ of one ground state of the pair differs by $\pi$ compared to the relative angle $(\theta)$ of other member of the pair. Specifically, if one ground state of the pair is characterized by $\theta = 0$ (for all $m$) then necessarily, the other member of the pair is characterized by $\theta = |\pi|$. The available sign-flipping transformation for the hopping term is the reason why the Hamiltonians are "colored" in this "ground-states" work only by the sign of the interaction-energy term.

Even the positive/negative-Hamiltonians are not independent of each other, since the ground state of one is (exactly) the highest energy state of the other and vice versa [52]. Or more generally – systems of vertical-mirror-imaged positions on the DNLSE phase diagram (resulting from sign-flipping the entire Hamiltonian) share the same statistical properties (*Figure 5*).

Generally, the DNLSE is not integrable and has no general (for arbitrary initial conditions) analytic solution. However, a set of analytic solutions in the form of discrete plane wave states does exist (Eq. **(13)**). All complex functions of a discrete plane wave state (numbered $q$) are of the same amplitude ($u_0$) and the phases of neighboring functions are all separated by the same discrete angle ($\alpha_q$). During evolution, all field functions acquire phase at the same rate. In terms of energy bands, the set of plane waves ($\{q\}$) correspond to the lowest energy band (see *optics* in section **9**). The two ground states of the positive-Hamiltonian pair are special plane wave "stationary states" where all amplitudes are real (with $\alpha_q = 0$ or $\alpha_q = \pi$) and all phases grow linearly with distance at a rate of $\Gamma \cdot u_0^2 \cdot z$ (Eq. **(15)**).

All four ground states – the two ground states of the positive-Hamiltonian pair as well as the two ground states of the negative-Hamiltonian pair satisfy a specific recurrence relations (Eq. **(22)**). The recurrence relations are found either through direct manipulations of the dynamic equations for the polar coordinates ($I_m, \phi_m$) (Eq. **(4)**), or through a specific variation of an expression that includes the two DNLSE conserved quantities **[21]**. The ground states of the positive-Hamiltonian pair are equal-amplitude solutions of the recurrence relations. The ground states of the negative-Hamiltonian pair are site-centered discrete breather solutions (*Figure 9* and the approximate analytic expressions of Eq. **(23)**).

Each of the two ground states of the systems of the positive DNLSE Hamiltonian pair, defined as "the smallest value of the Hamiltonian of all possible DNLSE solutions at a given density", is a zero entropy state (Eq. **(16)**), its DNLSE temperature is also zero (Eq. **(17)**) and the correlation distance ($k$) of the field functions of each of these two ground states extends to infinity (Eq. **(20)**).

In terms of stability, linear modulational instability analysis finds the high-energy discrete plane wave states of the positive-Hamiltonian unstable and finds the low-energy discrete plane wave states stable (cf. Eq. **(21)**). According to our suggested, more demanding, entropy-stability criterion, *all* discrete plane wave states except for the two ground-state plane waves are entropy-unstable under a location-preserving perturbation (*Figure 8*). Since such location-preserving perturbation cannot be applied to the ground states, we consider these ground states to be entropy-stable.

The abstract $1d$ DNLSE discussed in this work appears in the analysis of physical systems in several fields such as magnetism, optics and ultracold atoms. We elaborate on the specifics of the DNLSE in each of these fields with emphasis on ground states. In particular, following **[13]**, we show how the dynamics of a 1d optically-trapped BEC, in a rather wide range of system densities and system nonlinearities, can be described by the classical DNLSE as given by Eq. **(2)** with $signC = -1$ and $\Gamma > 0$.